\documentclass[11pt,fleqn]{article}
\usepackage{latexsym}
\usepackage{amssymb}
\usepackage{amsmath}
\usepackage{amscd}
\usepackage{amsthm}
\usepackage[left=2cm,top=2.5cm,right=2.5cm,bottom=2cm]{geometry}

\usepackage[dvips]{graphicx}
\usepackage{hyperref}
\begin{document}
\begin{center}
\Large{\bf{SOME FRW MODELS OF ACCELERATING UNIVERSE WITH DARK
ENERGY}}
\\
%\vspace{10mm} \normalsize{} \vspace{5mm}
\vspace{10mm} \normalsize{Suresh Kumar}\\ \vspace{5mm} \normalsize{
Department of Applied Mathematics,\\ Delhi Technological University
(Formerly Delhi College of Engineering),  \\Bawana Road, Delhi-110
042, India.} \normalsize{E-mail: sukuyd@gmail.com}
\end{center}
%%\date{}
%%\maketitle
\begin{abstract}
The paper deals  with a spatially homogeneous and isotropic FRW
space-time filled with perfect fluid and dark energy components. The
two sources are assumed to interact minimally, and therefore their
energy momentum tensors are conserved separately. A special law of
variation for the Hubble parameter proposed by Berman (1983) has
been utilized to solve the field equations. The Berman's law yields
two explicit forms of the scale factor governing the FRW space-time
and constant values of deceleration parameter. The role of dark
energy with variable equation of state parameter has been studied in
detail in the evolution of FRW universe. It has been found that dark
energy dominates the universe at the present epoch, which is
consistent with the observations. The physical behavior of the
universe is discussed in detail.
\end{abstract}
\smallskip


\vspace{1.5cm}
%%%%%%%%%%%%%%%%%%%%%%%%%%%%%%%%%%%%%%%%%%%%%%%%%%%%%%%%%%%%%%%%%%%%%%%%%%%%%%%%%%%%%%%%%%%%%%%%%%%%%
%%%%%%%%%%%%%%%%%%%%%%%%%%%%%%%   SECTION 1  %%%%%%%%%%%%%%%%%%%%%%%%%%%%%%%%%%%%%%%%%%%%%%%%%%%%%%%%%%
\section{Introduction}
Recent observations of type Ia supernovae (SN Ia) [1-5], galaxy
redshift surveys \cite{6}, cosmic microwave background radiation
(CMBR) data \cite{7,8} and large scale structure \cite{9} strongly
suggest that the observable universe is undergoing an accelerated
expansion. Observations also suggest that there had been a
transition of the universe from the earlier deceleration phase to
the recent acceleration phase \cite{10}. The cause of this sudden
transition and the source of the accelerated expansion is still
unknown. Measurements of CMBR anisotropies, most recently by the
WMAP satellite, indicate that the universe is very close to flat.
For a flat universe, its energy density must be equal to a certain
critical density, which demands a huge contribution from some
unknown energy stuff. Thus, the observational effects like the
cosmic acceleration, sudden transition, flatness of universe and
many more, need explanation. It is generally believed that some sort
of `dark energy' (DE) is pervading the whole universe. It is a
hypothetical form of energy that permeates all of space and tends to
increase the rate of expansion of the universe \cite{11}. The most
recent WMAP observations indicate that DE accounts for 72\% of the
total mass energy of the universe \cite{12}. However, the nature of
DE is still a mystery.

Many cosmologists believe that the simplest candidate for the DE is
the cosmological constant ($\Lambda$) or vacuum energy since it fits
the observational data well. During the cosmological evolution, the
$\Lambda$-term has the constant energy density and pressure
$p^{(de)}=-\rho^{(de)}$, where the superscript $(de)$ stands for DE.
However, one has the reason to dislike the cosmological constant
since it always suffers from the theoretical problems such as the
``fine-tuning" and ``cosmic coincidence" puzzles \cite{12a}. That is
why, the different forms of dynamically changing DE with an
effective equation of state (EoS),
$\omega^{(de)}=p^{(de)}/\rho^{(de)}<-1/3$, have been proposed in the
literature. Other possible forms of DE include quintessence
($\omega^{(de)}>-1$)\cite{13}, phantom ($\omega^{(de)}<-1$)
\cite{14} etc. While the possibility $\omega^{(de)}<<-1$ is ruled
out by current cosmological data from SN Ia (Supernovae Legacy
Survey, Gold sample of Hubble Space Telescope) \cite{5,15}, CMBR
(WMAP, BOOMERANG) \cite{16,17} and large scale structure (Sloan
Digital Sky Survey) \cite{18} data, the dynamically evolving DE
crossing the phantom divide line (PDL) $(\omega^{(de)}=-1)$ is
mildly favored. SN Ia data collaborated with CMBR anisotropy and
galaxy clustring statistics suggest that $-1.33<\omega^{(de)}<-0.79$
(see, Tegmark et al. \cite{18a}).

In 1983, Berman \cite{19} proposed a special law of variation of
Hubble parameter in FRW space-time, which yields a constant value of
deceleration parameter (DP). Such a law of variation for Hubble's
parameter is not inconsistent with the observations and is also
approximately valid for slowly time-varying DP models. The law
provides explicit forms of scale factors governing the  FRW universe
and facilitates to describe accelerating as well as decelerating
modes of evolution of the universe. Models with constant DP have
been extensively studied in the literature in different contexts
(see, Kumar and Singh \cite{20} and references therein). Most of the
models with constant DP have been studied by considering perfect
fluid or ordinary matter in the universe. But the ordinary matter is
not enough to describe the dynamics of an accelerating universe as
mentioned earlier. This motivates the researchers to consider the
models of the universe filled with some exotic type of matter such
as the DE along with the usual perfect fluid. Recently, some dark
energy models with constant DP have been investigated by Akarsu and
Kilinc \cite{21,21a,22}, Yadav \cite{22a} and Yadav and Yadav
\cite{22b}.

In this paper, we have considered minimally interacting perfect
fluid and DE energy components with constant DP within the framework
of a FRW space-time in general relativity. The paper is organized as
follows. In Sect. 2, the model and field equations have been
presented. Sect. 3 deals with the exact solutions of the field
equations and physical behavior of the model. Finally, concluding
remarks have been given in Sect. 4.

\section{Model and field equations}

In$\;$ standard spherical coordinates $\;(x^{i})=(t,r,\theta,\phi)$,
a spatially homogeneous and isotropic FRW line element has the form
(in units $c=1$)
\begin{equation}\label{1}
ds^{2}=-dt^{2}+a^{2}(t)\left[\frac{dr^{2}}{1-kr^{2}}+r^{2}(d\theta^{2}
+\text{sin}^{2}\theta\; d\phi^{2})\right],
\end{equation}
where $a(t)$ is the cosmic scale factor, which describes how the
distances (scales) change in an expanding or contracting universe,
and is related to the redshift of the 3-space; $k$ is the curvature
parameter, which describes the geometry of the spatial section of
space-time with closed, flat and open universes corresponding to
$k=-1,$\;$ 0,$\;$ 1\;$, respectively. The coordinates $r$, $\theta$
and $\phi$ in the metric \eqref{1} are `comoving' coordinates. The
FRW models have been remarkably successful in describing the
observed nature of universe.

The Einstein's field equations in case of a mixture of perfect fluid
and DE components, in the units $8\pi G=c=1$,  read as
\begin{equation}\label{2}
R_{ij}-\frac{1}{2} g_{ij}R =- T_{ij},
\end{equation}
where $T_{ij}=T^{(m)}_{ij}+T^{(de)}_{ij}$ is the overall energy
momentum tensor with $T^{(m)}_{ij}$ and $T^{(de)}_{ij}$ as the
energy momentum tensors of ordinary matter and DE, respectively.
These are given by
\begin{eqnarray}\label{3}
T^{(m)\;i}_{\;\;j}&=&\text{diag}\;[-\rho^{(m)},\;p^{(m)}\;,p^{(m)},\;p^{(m)}]\nonumber\\
                   &=&\text{diag}\;[-1,\;\omega^{(m)}\;,\omega^{(m)},\;\omega^{(m)}]\rho^{(m)}
\end{eqnarray}
and
\begin{eqnarray}\label{4}
T^{(de)\;i}_{\;\;j}&=&\text{diag}\;[-\rho^{(de)},\;p^{(de)}\;,p^{(de)},\;p^{(de)}]\nonumber\\
                   &=&\text{diag}\;[-1,\;\omega^{(de)}\;,\omega^{(de)},\;\omega^{(de)}]\rho^{(de)}
\end{eqnarray}
where $\rho^{(m)}$ and $p^{(m)}$ are, respectively the energy
density and pressure of the perfect fluid component or ordinary
baryonic matter while $\omega^{(m)}=p^{(m)}/\rho^{(m)}$ is its EoS
parameter. Similarly,  $\rho^{(de)}$ and $p^{(de)}$ are,
respectively the energy density and pressure of the DE component
while $\omega^{(de)}=p^{(de)}/\rho^{(de)}$ is the corresponding EoS
parameter.

In a comoving coordinate system, the field equations \eqref{2}, for
the FRW space-time \eqref{1}, in case of \eqref{3} and \eqref{4},
read as
\begin{equation}\label{2.5}
2\frac{\ddot{a}}{a}+\frac{\dot{a}^{2}}{a^{2}}+\frac{k}{a^{2}}=-\omega^{(m)}\rho^{(m)}-\omega^{(de)}\rho^{(de)},
\end{equation}
\begin{equation}\label{2.6}
3\frac{\dot{a}^{2}}{a^{2}}+3\frac{k}{a^{2}}
 =\rho^{(m)}+\rho^{(de)}.
\end{equation}

The energy conservation equation
$T^{(de)\;ij}_{\;\;\;\;\;\;\;\;\;\;;j} =0$ yields
\begin{equation}\label{2.11}
\dot\rho^{(m)}+3(1+\omega^{(m)})\rho^{(m)}H+\dot\rho^{(de)}+3(1+\omega^{(de)})\rho^{(de)}H=0,
\end{equation}
where $H=\dot{a}/a$ is the Hubble parameter.

\section{Solution of Field Equations}
The field equations \eqref{2.5} and \eqref{2.6} involve five unknown
variables, viz., $a$, $\omega^{(m)}$, $\omega^{(de)}$, $\rho^{(m)}$
and $\rho^{(de)}$. Therefore, to find a deterministic solution of
the equations, we need three suitable assumptions connecting the
unknown variables.

First, we assume that the perfect fluid and DE components interact
minimally. Therefore, the energy momentum tensors of the two sources
may be conserved separately.

The energy conservation equation
$T^{(m)\;ij}_{\;\;\;\;\;\;\;\;\;\;;j} =0$, of the perfect fluid
leads to
\begin{equation}\label{2.12}
\dot\rho^{(m)}+3(1+\omega^{(m)})\rho^{(m)}H=0,
\end{equation}
whereas the energy conservation equation
$T^{(de)\;ij}_{\;\;\;\;\;\;\;\;\;\;;j} =0$, of the DE component
yields
\begin{equation}\label{2.13}
\dot\rho^{(de)}+3(1+\omega^{(de)})\rho^{(de)}H=0.
\end{equation}

Next, we assume that the EoS parameter of the perfect fluid to be a
constant, that is,
\begin{equation}\label{14}
\omega^{(m)}=\frac{p^{(m)}}{\rho^{(m)}}=\text{const.},
\end{equation}
while $\omega^{(de)}$ has been allowed to be a function of time
since the current cosmological data from SN Ia, CMBR and large scale
structures mildly favor dynamically evolving DE crossing the PDL as
discussed in Section 1.

Now integration of \eqref{2.12} leads to
\begin{equation}\label{14a}
\rho^{(m)}=c_{0}a^{-3(1+\omega^{(m)})},
\end{equation}
where $c_{0}$ is a positive constant of integration.

Finally, we constrain the system of equations with a law of
variation for the Hubble parameter proposed by Berman \cite{19},
which yields a constant value of DP. The law reads as
\begin{equation}\label{15}
H=Da^{-n},
\end{equation}
where  $D>0$  and  $n\geq 0$ are constants.  In the following
subsections, we discuss the DE cosmology for $n\neq0$ and $n=0$ by
using the law \eqref{15}.

\subsection{DE Cosmology for $n\neq 0$}
In this case, integration of \eqref{15} leads to
\begin{equation}\label{16}
a(t)=(nDt+c_{1})^{\frac{1}{n}},
\end{equation}
where $c_{1}$ is a constant of integration. Therefore, the model
\eqref{1} becomes
\begin{equation}\label{19}
ds^{2}=-dt^{2}+(nDt+c_{1})^{\frac{2}{n}}\left[\frac{dr^{2}}{1-kr^{2}}+r^{2}(d\theta^{2}+\text{sin}^{2}\theta
d\phi^{2})\right].
\end{equation}

The Hubble parameter ($H$), energy density ($\rho^{(m)}$) of perfect
fluid, DE density ($\rho^{(de)}$) and EoS parameter
($\omega^{(de)}$) of DE, for the model \eqref{19} are found to be
\begin{equation}\label{20}
H=D(nDt+c_{1})^{-1},
\end{equation}
\begin{equation}\label{21}
\rho^{(m)}=c_{0}(nDt+c_{1})^{\frac{-3(1+\omega^{(m)})}{n}},
\end{equation}
\begin{equation}\label{22}
\rho^{(de)}=3D^{2}(nDt+c_{1})^{-2}+3k(nDt+c_{1})^{\frac{-2}{n}}
-c_{0}(nDt+c_{1})^{\frac{-3(1+\omega^{(m)})}{n}},
\end{equation}
\begin{equation}\label{23}
\omega^{(de)}=\frac{1}{\rho^{(de)}}\left[(2n-3)D^{2}(nDt+c_{1})^{-2}-k(nDt+c_{1})^{\frac{-2}{n}}
-c_{0}\omega^{(m)}(nDt+c_{1})^{\frac{-3(1+\omega^{(m)})}{n}}\right],
\end{equation}

The above solutions satisfy the equation \eqref{2.13} identically,
as expected.

The spatial volume ($V$) and expansion scalar $(\theta)$ of the
model read as
\begin{equation}\label{24}
V=a^{3}=(nDt+c_{1})^{\frac{3}{n}},
\end{equation}
\begin{equation}\label{25}
\theta=3H=3(nDt+c_{1})^{-1}.
\end{equation}

The density parameter $\Omega^{(m)}$ of perfect fluid and the
density parameter $\Omega^{(de)}$ of DE are given by
\begin{equation}\label{26}
\Omega^{(m)}=\frac{\rho^{(m)}}{3H^{2}}=\frac{c_{0}}{3D^{2}}(nDt+c_{1})^{\frac{2n-3(1+\omega^{(m)})}{n}},
\end{equation}
\begin{equation}\label{27}
\Omega^{(de)}=\frac{\rho^{(de)}}{3H^{2}}=1+\frac{3k}{D^{2}}(nDt+c_{1})^{\frac{-2(1-n)}{n}}
-\frac{c_{0}}{3D^{2}}(nDt+c_{1})^{\frac{2n-3(1+\omega^{(m)})}{n}}.
\end{equation}

The value of DP ($q$) is found to be
\begin{equation}\label{28}
q=-\frac{a\ddot{a}}{a^{2}}=n-1,
\end{equation}
which is a constant. The sign of $q$ indicates whether the model
inflates or not. A positive sign of $q$, i.e., $n>1$ corresponds to
the standard decelerating model whereas the negative sign of $q$,
i.e., $0< n<1$ indicates inflation. The expansion of the universe at
a constant rate corresponds to $n=1$, that is, $q=0$. Also, recent
observations of SN Ia [1-5, 28-30] reveal that the present universe
is accelerating and value of DP lies somewhere in the range
$-1<q<0.$ It follows that in the derived model, one can choose the
values of DP consistent with the observations.

We observe that at $t=-c_{1}/nD$, the spatial volume vanishes while
all the parameters diverge. Therefore, the model has a big bang
singularity at $t=-c_{1}/nD$, which can be shifted to $t=0$ by
choosing $c_{1}=0$. The cosmological evolution in FRW space-time is
expansionary, since the scale factor monotonically increases with
the cosmic time. So, the universe starts expanding with a big bang
singularity in the derived model. The parameters $H$, $\rho^{(m)}$,
$\rho^{(de)}$ and $\theta$ start off with extremely large values,
and continue to decrease with the expansion of the universe. The
spatial volume grows with the cosmic time.

\begin{figure}[h]
\begin{center}
\includegraphics[height=7.1275 cm]{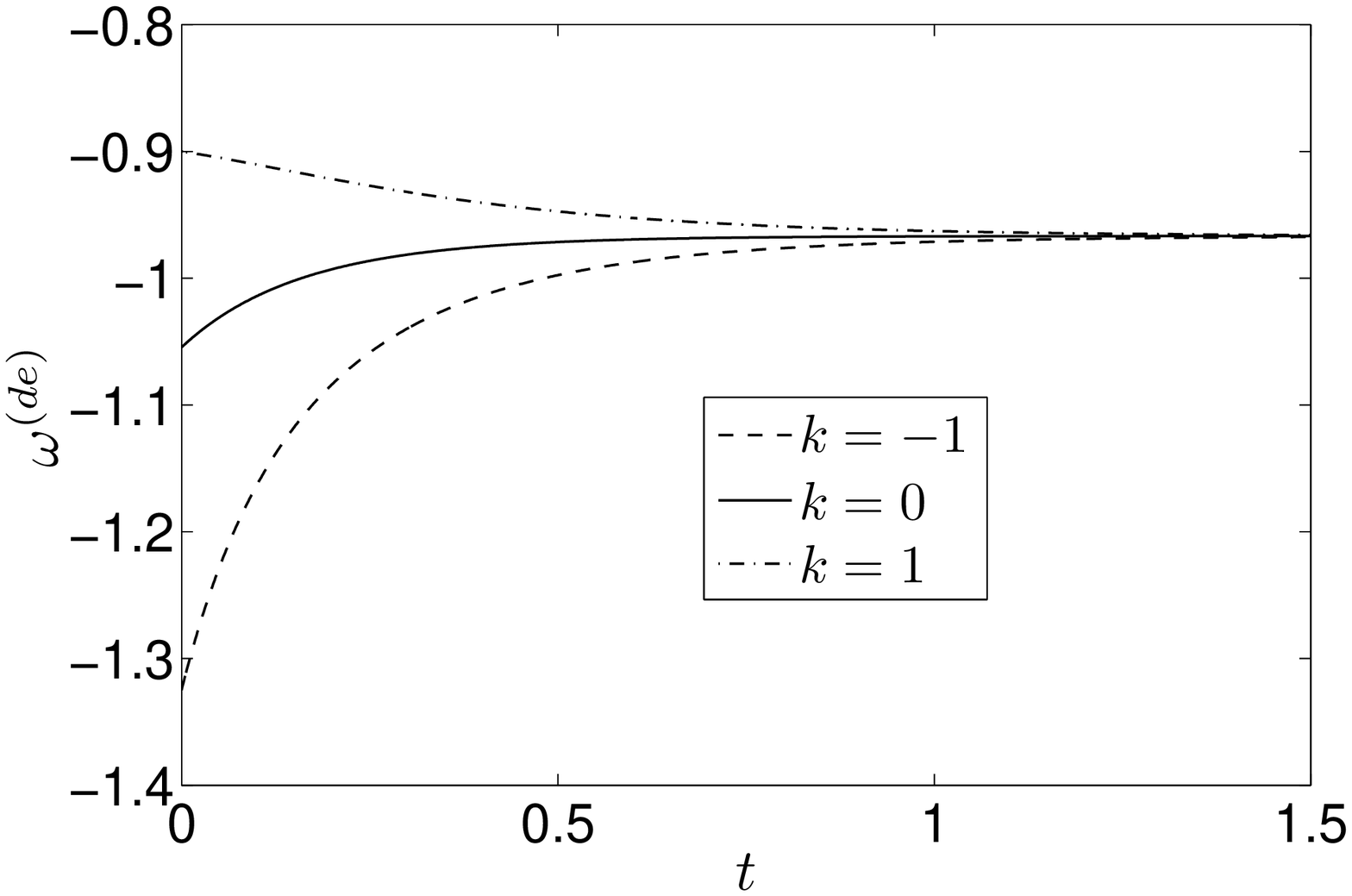}
\caption{\textsf{$\omega^{(de)}$ versus $t$ with $n=0.05$, $D=2$,
$c_{0}=c_{1}=1$, $\omega^{(m)}=0$.}}
\end {center}
\end{figure}

\begin{figure}[h]
\begin{center}
\includegraphics[height=7.1275 cm]{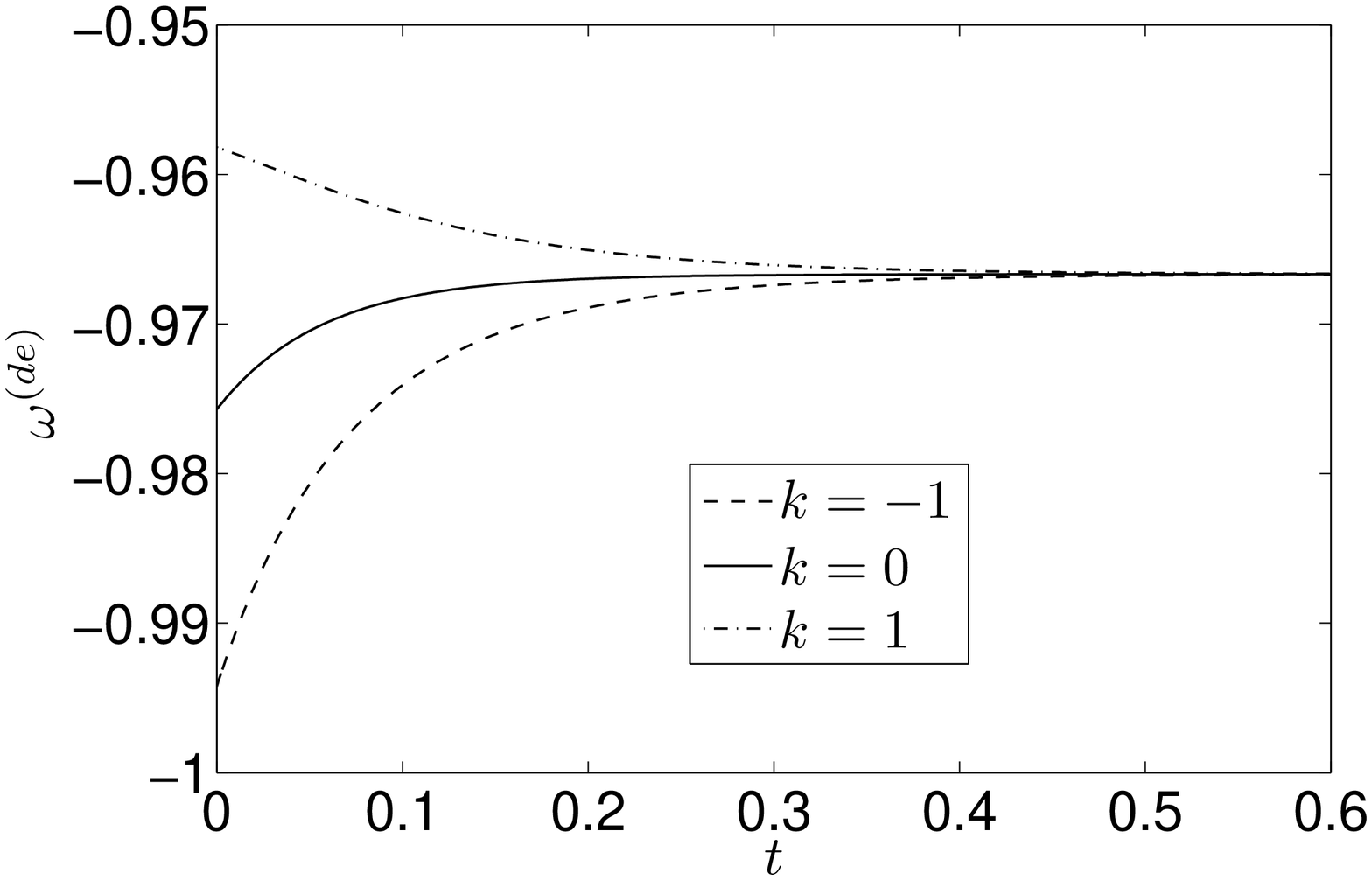}
\caption{\textsf{$\omega^{(de)}$ versus $t$ with $n=0.05$, $D=6$,
$c_{0}=c_{1}=1$, $\omega^{(m)}=0$.}}
\end {center}
\end{figure}

The EoS parameter $\omega^{(de)}$ of DE asymptotically approaches to
$\frac{2n}{3}-1$ provided $n<1$. Thus, the dynamics of
$\omega^{(de)}$ depends on $n$ for sufficiently large values of $t$.
Fig.1 and Fig. 2 show the variation of $\omega^{(de)}$ during the
cosmic evolution of closed ($k=-1$), flat ($k=0$) and open ($k=1$)
universes for $D=2$ and $D=6$, respectively. From Fig.1, we observe
that for closed and flat universes, $\omega^{(de)}$ starts in
phantom region $(\omega^{(de)}<-1)$, crosses the PDL
$(\omega^{(de)}=-1)$ and finally varies in the quintessence region
$(\omega^{(de)}>-1)$. However, it evolves within the quintessence
region only for open universe. For $D=6$ (see Fig. 2),
$\omega^{(de)}$ varies within the quintessence region throughout the
evolution of the three universes. Thus, the nature of DE depends on
the constants involved in the expression \eqref{23} of
$\omega^{(de)}$. Also, observations predict that
$-1.33<\omega^{(de)}<-0.79$ (see, Tegmark et al. \cite{18a}).
Therefore, Fig. 1 and Fig. 2 suggest that the derived model is
consistent with the observations.

Adding \eqref{26} and \eqref{27}, we get the overall density
parameter
\begin{equation}\label{29}
\Omega=\Omega^{(m)}+\Omega^{(de)}=1+\frac{3k}{D^{2}}(nDt+c_{1})^{\frac{-2(1-n)}{n}}.
\end{equation}

This shows that $\Omega<1$, $\Omega=1$ and $\Omega>1$ according as
$k=-1$ (closed universe),  $k=0$ (flat universe) and $k=1$ (open
universe) respectively, as expected. We find that $\Omega\approx1$
in the cases $k=-1$ and $k=1$ for sufficiently large values of $t$
provided $n<1$. Thus, the model predicts a flat universe for large
times. Since the present-day universe is very close to flat, so the
derived model is consistent with the observations.

\begin{figure}[h]
\begin{center}
\includegraphics[height=7.1275 cm]{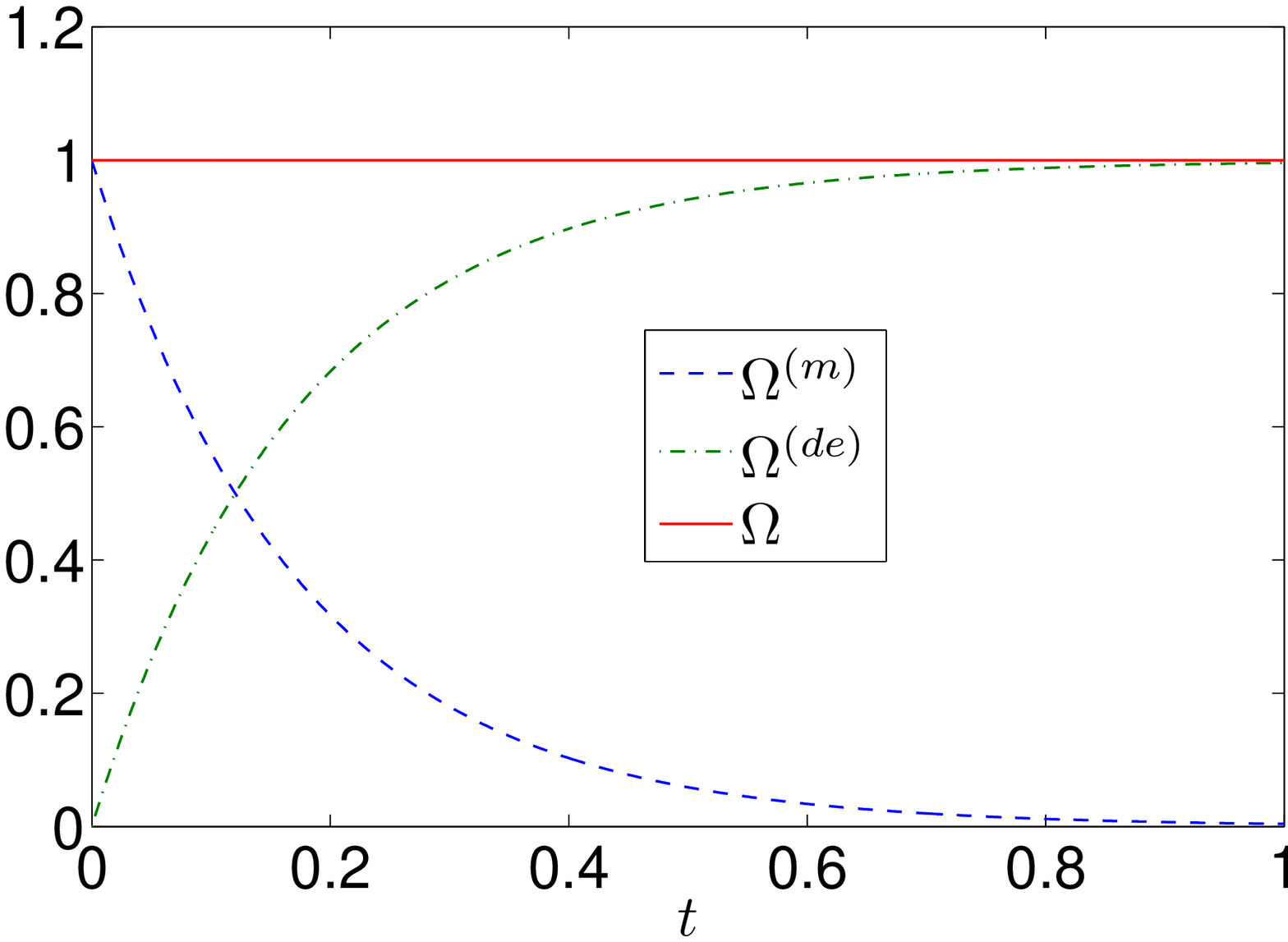}
\caption{\textsf{Density parameters versus $t$ with $n=0.05$, $D=2$,
$c_{0}=12$, $c_{1}=1$, $\omega^{(m)}=0$.}}
\end {center}
\end{figure}

Fig. 3 demonstrates the behavior of density parameters in the
evolution of a flat universe ($k=0$). We observe that initially the
ordinary matter density dominates the universe. But later on, the DE
dominates the evolution, which is probably the possible cause of
acceleration of the present universe.

\subsection{DE Cosmology for $n=0$}
In this case, integration of \eqref{15} yields
\begin{equation}\label{17}
a(t)=c_{2}e^{Dt},
\end{equation}
where $c_{2}$ is a positive constant of integration. Therefore, the
model \eqref{1} becomes
\begin{equation}\label{19a}
ds^{2}=-dt^{2}+c_{2}^{2}e^{2Dt}\left[\frac{dr^{2}}{1-kr^{2}}+r^{2}(d\theta^{2}+\text{sin}^{2}\theta
d\phi^{2})\right].
\end{equation}

The Hubble parameter, energy density of perfect fluid, DE density
and EoS parameter of DE, for the model \eqref{19a} are obtained as
\begin{equation}\label{20}
H=D,
\end{equation}
\begin{equation}\label{21}
\rho^{(m)}=c_{0}c_{2}^{-3(1+\omega^{(m)})}e^{-3D(1+\omega^{(m)})t},
\end{equation}
\begin{equation}\label{22}
\rho^{(de)}=3D^{2}+3kc_{2}^{-2}e^{-2Dt}
-c_{0}c_{2}^{-3(1+\omega^{(m)})}e^{-3D(1+\omega^{(m)})t},
\end{equation}
\begin{equation}\label{22}
\omega^{(de)}=\frac{1}{\rho^{(de)}}\left[-3D^{2}-kc_{2}^{-2}e^{-2Dt}-
c_{0}c_{2}^{-3(1+\omega^{(m)})}\omega^{(m)}e^{-3D(1+\omega^{(m)})t}\right].
\end{equation}

The above solutions satisfy the equation \eqref{2.13} identically,
as expected.

The spatial volume and expansion scalar of the model read as
\begin{equation}\label{30}
V=c_{2}^{3}e^{3Dt},
\end{equation}
\begin{equation}\label{30}
\theta=3D.
\end{equation}

The density parameters of perfect fluid and DE are given by
\begin{equation}\label{18}
\Omega^{(m)}=\frac{c_{0}c_{2}^{-3(1+\omega^{(m)})}}{3D^{2}}e^{-3D(1+\omega^{(m)})t},
\end{equation}
\begin{equation}\label{18}
\Omega^{(de)}=1+\frac{3kc_{2}^{-2}}{3D^{2}}e^{-2Dt}
-\frac{c_{0}c_{2}^{-3(1+\omega^{(m)})}}{3D^{2}}e^{-3D(1+\omega^{(m)})t}.
\end{equation}

The DP is given by
\begin{equation}\label{18}
q=-1.
\end{equation}

Recent observations of SN Ia [1-5, 28-30] suggest that the universe
is accelerating in its present state of evolution. It is believed
that the way universe is accelerating presently; it will expand at
the fastest possible rate in future and forever. For  $n=0$, we get
\textbf{ $q=-1$ }; incidentally this value of DP leads to $dH/dt=0$,
which implies the greatest value of Hubble's parameter and the
fastest rate of expansion of the universe. Therefore, the derived
model can be utilized to describe the dynamics of the late time
evolution of the actual universe. So, in what follows, we emphasize
upon the late time behavior of the derived model.

\pagebreak

\begin{figure}[h]
\begin{center}
\includegraphics[height=7.1275 cm]{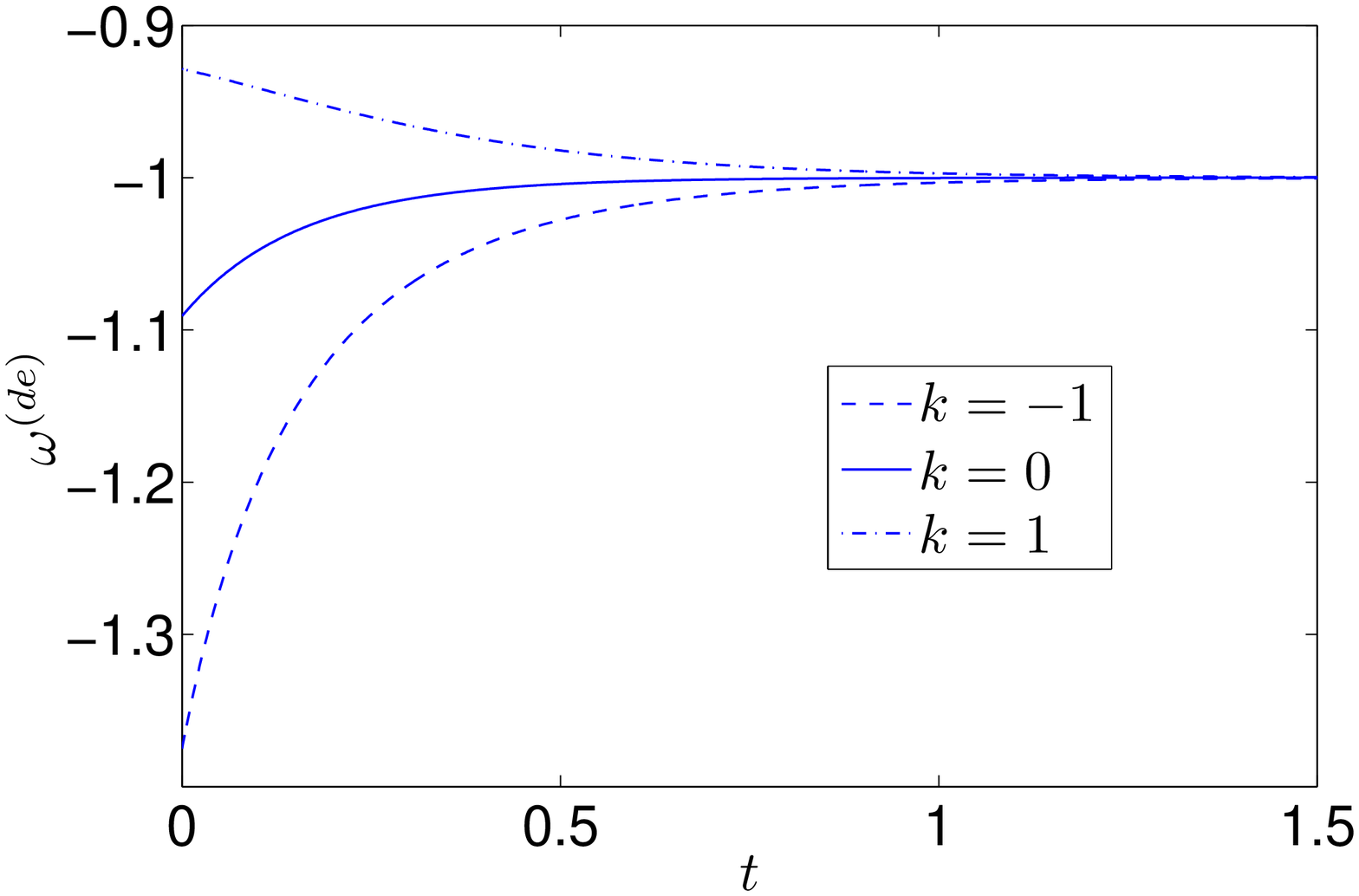}
\caption{\textsf{$\omega^{(de)}$ versus $t$ with $D=2$,
$c_{0}=c_{2}=1$, $\omega^{(m)}=0$.}}
\end {center}
\end{figure}

\begin{figure}[h]
\begin{center}
\includegraphics[height=7.1275 cm]{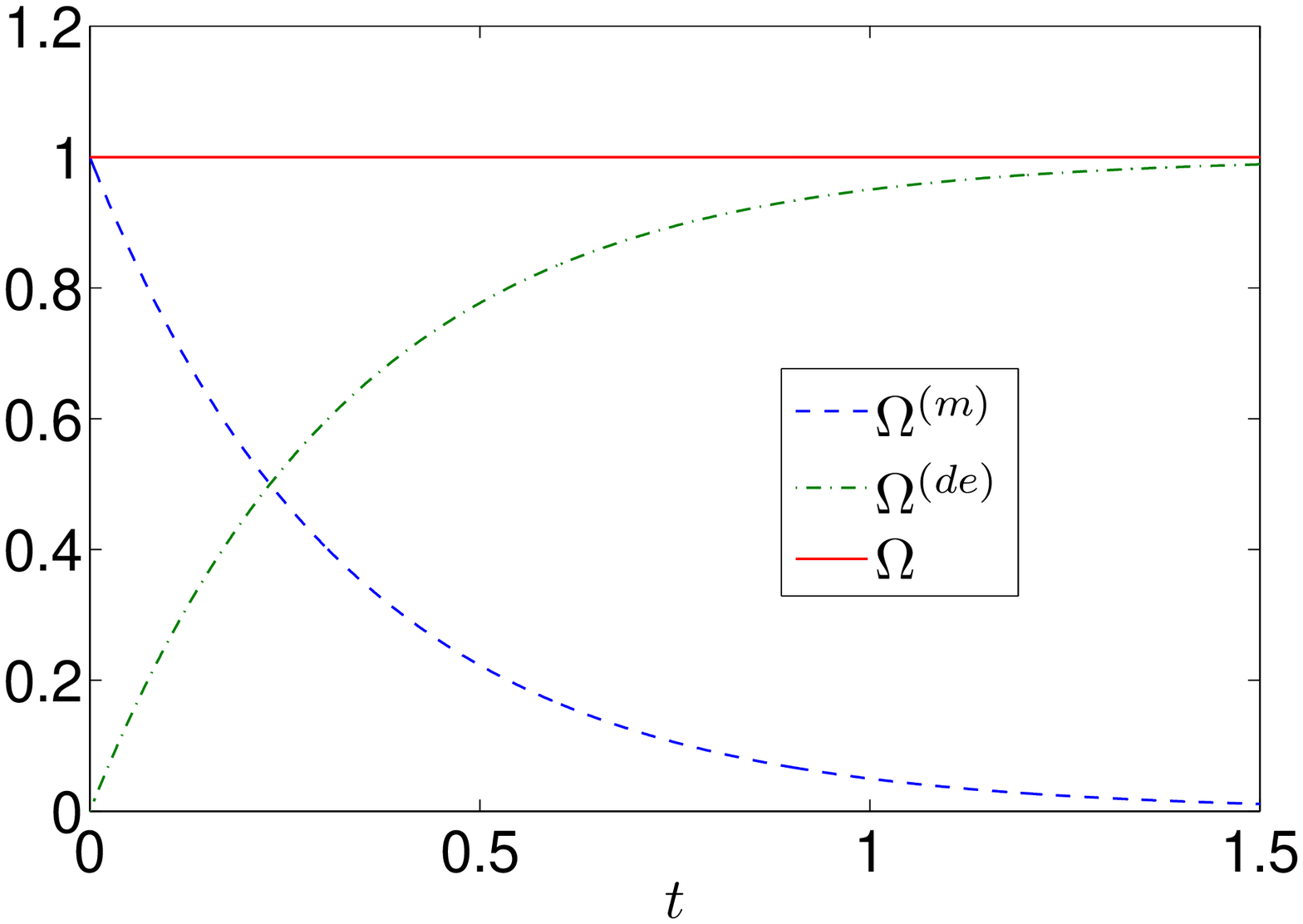}
\caption{\textsf{Density parameters versus $t$ with $D=1$,
$c_{0}=3$, $c_{2}=1$, $\omega^{(m)}=0$.}}
\end {center}
\end{figure}

 Fig. 4 depicts the behavior of the EoS parameter
$\omega^{(de)}$ of DE in the closed, flat and open universes. We
observe that for sufficiently large time, $\omega^{(de)}\approx -1$
in each case. Therefore, the so called cosmological constant is a
suitable candidate to represent the behavior of DE in the derived
model at late times.

Further, at late times, we have

\[\rho^{(m)}\approx0\]
and
\[\rho^{(de)}\approx3D^{2}.\]

This shows that the ordinary matter density becomes negligible
whereas the accelerated expansion of the universe continues with
non-zero and constant DE density at late times, as predicted by the
observations. Fig. 5 illustrates the behavior of the density
parameters during the evolution of the universe in the derived
model. It is observed that the DE component dominates the universe
at late times and $\Omega=\Omega^{(m)}+\Omega^{(de)}\approx1.$

\section{Concluding Remarks}
The special law of variation for Hubble's parameter proposed by
Berman in FRW space-time yields constant value of DP given by
$q=n-1$, which provides accelerating models of the universe for
$n<1$ and decelerating ones for $n>1$. In the present work, we have
emphasized that decelerating models can be described by the usual
perfect fluid, while dynamics of accelerating universe can be
described by considering some exotic type of matter such as the DE.
We have investigated the role of DE with variable EoS parameter in
the evolution of universe within the framework of a spatially
homogeneous FRW space-time by taking into account the special law of
variation of Hubble parameter. DE cosmologies have been discussed in
two different cases, viz., power-law ($n\neq 0$) and exponential-law
($n=0$). The analysis of the models reveals that the present-day
universe is dominated by DE, which can successfully describe the
accelerating nature of the universe consistent with the
observations.

\end{document}